\documentclass[sn-mathphys-ay, iicol]{sn-jnl}

\usepackage{graphicx}%
\usepackage{multirow}%
\usepackage{amsmath,amssymb,amsfonts,mathtools}%
\usepackage{amsthm}%
\usepackage{mathrsfs}%
\usepackage[title]{appendix}%
\usepackage{xcolor}%
\usepackage{textcomp}%
\usepackage{manyfoot}%
\usepackage{booktabs}%
\usepackage{algorithm}%
\usepackage{algorithmicx}%
\usepackage{algpseudocode}%
\usepackage{listings}%
\usepackage{cuted} 
\usepackage{mathpazo}
\usepackage{enumitem}
\usepackage{array, longtable, calc, hhline, xfrac, comment, subcaption, hyperref}
\usepackage{enumitem}
\newlist{notelist}{description}{1}
\setlist[notelist]{labelwidth=3.2em,leftmargin=!,style=nextline}

\newtheorem{proposition}{Proposition}
\newtheorem{theorem}{Theorem}
\newtheorem{corollary}{Corollary}
\newtheorem{lemma}{Lemma}

\def\A{\mathbf A}
\def\B{\mathbf B}
\def\C{\mathbf C}
\def\G{\mathbf G}
\def\H{\mathbf H}
\def\I{\mathbf I}
\def\M{\mathbf M}
\def\N{\mathbf N}
\def\U{\mathbf U}

\def\X{\mathcal X}
\def\supp{\mathcal S}

\def\0{\mathbf 0}
\def\u{\mathbf u}
\def\f{\mathbf f}
\def\s{\mathbf s}
\def\v{\mathbf v}

\newcommand{\e}{\mathbf e}
\newcommand{\w}{\mathbf w}
\newcommand{\x}{\mathbf x}
\newcommand{\z}{\mathbf z}

\def\RR{\mathbb R}

\def\eff{\mathrm{eff}}

\DeclareMathOperator{\Var}{Var}
\DeclareMathOperator{\cov}{cov}
\DeclareMathOperator{\tr}{tr}
\DeclareMathOperator{\diag}{diag}
\DeclareMathOperator{\rank}{rank}

\begin{document}

\title[Removing Redundant Candidates in Exact D-Optimal Design]{Removal of Redundant Candidate Points for the Exact D-Optimal Design Problem}


\author*[1]{\fnm{Radoslav} \sur{Harman}}\email{radoslav.harman@fmph.uniba.sk}
\author[1]{\fnm{Samuel} \sur{Rosa}}

\affil*[1]{\orgdiv{Faculty of Mathematics, Physics and Informatics}, \orgname{Comenius University}, \orgaddress{\street{Mlynsk\'{a} Dolina F1}, \city{Bratislava}, \postcode{842 48}, \country{Slovakia}}}


\abstract{
One of the most common problems in statistical experimentation is computing D-optimal designs on large finite candidate sets. While optimal approximate (i.e., infinite-sample) designs can be efficiently computed using convex methods, constructing optimal exact (i.e., finite-sample) designs is a substantially more difficult integer-optimization problem. In this paper, we propose necessary conditions, based on approximate designs, that must be satisfied by any support point of a D-optimal exact design. These conditions enable rapid elimination of redundant candidate points without loss of optimality, thereby reducing memory requirements and runtime of subsequent exact design algorithms. In addition, we prove that for sufficiently large sample sizes, the supports of D-optimal exact designs are contained in a typically small maximum-variance set. We demonstrate the approach on randomly generated benchmark models with candidate sets up to 100 million points, and on commonly used constrained mixture models with up to one million points. The proposed approach reduces the initial candidate sets by several orders of magnitude, thereby making it possible to compute exact D-optimal designs for these problems via mixed-integer second-order cone programming, which provides optimality guarantees.
}



\maketitle


\section{Introduction}

\subsection{Optimal design of experiments}

An experiment is commonly understood as a set of $n$ trials, each yielding an observation. The field of optimal experimental design seeks to select these trials so as to maximize information obtained from the experiment as a whole (see, e.g., the monographs \cite{fedorov1972}, \cite{pazman86}, \cite{pukelsheim}, \cite{AtkinsonEA07}, \cite{GoosJones}, \cite{pronzatopazman2013}, and \cite{lopezfidalgo2023}).

A central component of any experiment is a set $\X$ that represents the conditions under which individual trials may be conducted. We call the elements of $\X$ candidate points or candidates. Often, they correspond to all permissible combinations of the levels of several controllable factors. In this paper, we assume that $\X$ contains a finite number $N \geq 2$ of candidates: $\X=\{\x_1,\ldots,\x_N\}$. While a finite $\X$ is usually the only realistic or meaningful choice, its size $N$ can be very large. 

We denote the required number of trials by $n$ and formally represent a size-$n$ design by a nonnegative vector $\w = (w_1, \ldots, w_N)^\top$, such that $\sum_{i=1}^N w_i = 1$ and $nw_1,\ldots,nw_N$ are nonnegative integers. Here, $n\w$ is the vector of replications, i.e., its components are the numbers of trials performed at $\x_1,\ldots,\x_N$. The set of all such size-$n$ designs $\w$ is the set $\Xi_n$ of lattice points of the probability simplex in $\RR^N$, consisting of $\binom{N+n-1}{N-1}$ vectors whose components are integer multiples of $1/n$.

In the field of optimal experimental design, it is also common to consider the full probability simplex in $\RR^N$, which we denote by $\Xi_\infty$. A vector $\w \in \Xi_\infty$ represents the proportions of trials at $\x_1,\ldots,\x_N$, without any integrality constraint. Since the real-valued proportions specified by $\w$ can, in general, be realized only as the number of trials tends to infinity, such designs are referred to as approximate. In contrast, the size-$n$ designs in $\Xi_n$ are called exact. Under our formalization we have $\Xi_n \subseteq \Xi_\infty$; therefore, if we use the term design without any qualifier, we cover both approximate and exact designs. An important set associated with a design $\w$ is its support
\begin{equation*}
    \supp(\w) := \{i \in [N] : w_i > 0\}, 
\end{equation*}
where $[N]:=\{1,\ldots,N\}$. Each $\ell \in \supp(\w)$ is called a support index of $\w$. 

The quality of a design $\w$, with respect to its ability to provide precise estimates of the parameters of an underlying statistical model, is measured via its information matrix. The most common form of the information matrix is
\begin{equation*}
  \M(\w):=\sum_{i=1}^N w_i \f_i\f_i^\top,
\end{equation*}
where $\f_1,\ldots,\f_N \in \RR^m$ are known vectors. This matrix typically represents a normalized version of the Fisher information matrix for the $m$-dimensional model parameter, and its additive form arises from the assumption that the observations are independent across trials.

The vectors $\f_1, \ldots, \f_N$ are based on the model; for instance, for the most common case, i.e., the linear regression with uncorrelated and homoscedastic observations $y$ having expectation $\f^\top(\x)\boldsymbol{\theta}$ for any $\x \in \X$, we simply have $\f_i=\f(\x_i)$ for all $i \in [N]$. However, the same form of the information matrix also appears in the locally optimal design approach for nonlinear models (e.g., Chapter 3 of \cite{lopezfidalgo2023}) and for generalized linear models (e.g., \cite{khuri}). Also, it covers heteroscedastic models via standard transformations (see, e.g., Remark 2.1 in \cite{lopezfidalgo2023}). In summary, the considered form of the information matrix is relevant for the majority of real applications.

Throughout, we restrict attention to settings in which $\M(\w)$ is nonsingular for some $\w \in \Xi_n$. This assumption ensures that there exists at least one experimental design for which the model parameters are estimable. It also implies $n \geq m$. Any design with a nonsingular information matrix is called nonsingular. For simplicity, we confine our attention to the nontrivial case $m \geq 2$.

To compare designs based on the size of their information matrices, we consider the criterion $\Phi$ of $D$-optimality, defined by
\begin{equation*}
    \Phi(\M) := \left(\det(\M)\right)^{1/m}
\end{equation*}
for all positive semidefinite $m \times m$ matrices $\M$. For the exact design problem, $D$-optimality was introduced by Wald in \cite{Wald} and has since served as a cornerstone of optimal design applications, due to its unique statistical and analytic properties (e.g., Section 6.2 of \cite{pukelsheim} or Chapter 11 of \cite{AtkinsonEA07}). Statistically, maximizing $\Phi$ is equivalent to minimizing the generalized variance of the parameter estimates, or the volume of the confidence ellipsoid for the unknown model parameter. Analytically, $\Phi$ is continuous, concave, and positively homogeneous on the set of all positive semidefinite matrices, as well as differentiable and strictly log-concave on the set of all positive definite $m \times m$ matrices.

An approximate design $\w^*_\infty$ is $D$-optimal if it maximizes $\Phi(\M(\w_\infty))$ over all $\w_\infty \in \Xi_\infty$; the set of all such designs is denoted by $\Xi^*_\infty$. Likewise, an exact design $\w_n^* \in \Xi_n$ is a $D$-optimal size-$n$ design if it maximizes $\Phi(\M(\w_n))$ over all $\w_n \in \Xi_n$. We denote the set of $D$-optimal size-$n$ designs by $\Xi^*_n$. Since $D$-optimality is the only criterion considered in this paper, we omit the prefix “$D$-” and simply refer to an optimal design.

Our assumptions ensure the existence of at least one optimal approximate design and at least one optimal size-$n$ design, and that the information matrices of these optimal designs are nonsingular. However, there may be infinitely many optimal approximate designs (see, e.g., \cite{HFR24}) and multiple optimal size-$n$ designs. The strict log-concavity of $\Phi$ on the set of positive definite matrices implies that all optimal approximate designs $\w^*_\infty$ have the same information matrix, which we denote $\M_*$. In contrast, different optimal size-$n$ designs may have distinct information matrices.

For any nonsingular design $\w$ we define
\begin{equation*}
    v_i(\w):=\f_i^\top \M^{-1}(\w) \f_i, \: i \in [N].
\end{equation*}
Viewed as a function of $i$, $v_i(\w)$ is sometimes called the variance function, because for a nonsingular size-$n$ design $\w_n$, the quantity $(\sigma^2/n) \cdot v_i(\w_n)$ equals the variance of the best linear unbiased estimator $\widehat{\f_i^\top\boldsymbol{\theta}}$ of the mean response $\f^\top_i \boldsymbol{\theta}$ in the standard linear regression discussed above, in which $\Var(y) \equiv \sigma^2$. Clearly, if two designs have the same information matrix, their variance functions are the same. We denote the variance function common to all optimal approximate designs $\w^*_\infty$ as
\begin{equation*}
    v_i^*:=\f_i^\top \M^{-1}(\w^*_\infty) \f_i, \: i \in [N].
\end{equation*}
The variance function is key for the fundamental equivalence theorem for $D$-optimality (e.g., Section 7 of \cite{pukelsheim}): 
\begin{proposition}[Equivalence theorem]\label{prop:eqt}
A nonsingular design $\w_\infty$ is an optimal approximate design if and only if
\begin{equation*}
    \max_{i \in [N]} v_i(\w_\infty) = m.
\end{equation*}
If $\w^*_\infty$ is an optimal approximate design, then $\supp(\w^*_\infty)$ is a subset of
\begin{equation*}
 \tilde{\supp}_\infty:= \{i \in [N] : v^*_i = m\}.
\end{equation*}
\end{proposition}
The maximum-variance set $\tilde{\supp}_\infty$ from Proposition \ref{prop:eqt} plays an important role in this paper. Note that even if the optimal approximate design is not unique, $\tilde{\supp}_\infty$ is. By the equivalence theorem, $\tilde{\supp}_\infty$ covers the set \begin{equation*}
    \supp^*_\infty := \cup_{\w^*_\infty \in \Xi^*_\infty} \supp(\w^*_\infty),
\end{equation*}
as a subset, i.e., $\tilde{\supp}_\infty$ contains any support point of any optimal approximate design. In most settings, the two sets actually coincide: $\tilde{\supp}_\infty = \supp^*_\infty$. 

In certain specific cases $\tilde{\supp}_\infty=[N]$. This occurs, for example, in some basic two-level factor models with no restrictions on factor combinations (e.g., the binary-level variants of the models 3.2, 3.3, and 3.4 in \cite{HFR24}), and in trigonometric regression on a regular grid over $[0,2\pi]$ (see Section 9.16 of \cite{pukelsheim}). Nevertheless, crucially for this paper, the size of $\tilde{\supp}_\infty$ is typically much smaller than $N$. An intuitive explanation follows from a geometric implication of the equivalence theorem: $v_\ell^*=m$ for all $\ell \in \tilde{\supp}_\infty$ means that all vectors $\f_\ell$, $\ell \in \tilde{\supp}_\infty$, lie exactly on the boundary of a common ellipsoid. In fact, this ellipsoid is the minimum-volume enclosing ellipsoid (MVEE) covering the vectors $\pm \f_1, \ldots, \pm \f_N$ (e.g., Section 2 of \cite{DBIG}). Note, however, that this simple boundary property is specific to optimal approximate designs; for the exact design problem, some support indices $\ell$ may correspond to regressors $\f_\ell$ lying strictly inside the MVEE.

For positively homogeneous criteria, such as $\Phi$, the relative efficiency of a design $\w$ and a nonsingular design $\tilde{\w}$ is defined as the ratio of the criterion values of their information matrices. In particular, the relative efficiency of $\w$ with respect to an optimal approximate design is
\begin{equation*}
    \eff_\infty(\w):=\frac{\Phi(\M(\w))}{\Phi(\M_*)}.
\end{equation*}
This quantity also provides a lower bound on the efficiency of $\w$ relative to any design.

The analytic forms of optimal size‑$n$ designs are generally complex and known only for special cases (e.g., \cite{Gaffke1986}, \cite{Neubauer2000}, \cite{Bailey2009}), and the computation of provably optimal exact designs is fundamentally challenging (\cite{Welch1982}). For small- and medium-sized problems, integer programming solvers can be used to construct verifiably optimal exact designs (\cite{sagnol2015}, \cite{Duarte2020}, \cite{Ahipasaoglu2021}). For larger problems, it is more common to rely on various heuristics, which typically yield designs with practically acceptable efficiency relative to the theoretical optimum; see, e.g., Chapter 12 of \cite{AtkinsonEA07} for a survey of the point exchange methods, or Subsection 2.3.9 and other subsections of \cite{GoosJones} for various coordinate exchange methods. 

In contrast, the problem of optimal approximate design (with a finite $\X$) is convex and therefore much easier to solve, both theoretically and computationally. Analytic forms of optimal approximate designs are known for many standard models (e.g., \cite{pukelsheim}) and, even when a closed-form solution is unavailable, an optimal approximate design can be constructed algorithmically to essentially perfect numerical precision for most practical problems with up to several million candidates (e.g., \cite{Yu2011}, \cite{ybt}, \cite{rex}).

For the application of some results in this paper, we therefore assume that an optimal approximate design $\w^*_\infty$ is known. Such a design is broadly useful for the optimal exact design problem in any case, as it can, for example:
\begin{itemize}
    \item provide often strong lower bounds on the efficiency of any size-$n$ design $\w_n$ relative to an optimal size-$n$ design $\w^*_n$: $\Phi(\M(\w_n))/\Phi(\M(\w^*_n)) \geq \eff_\infty(\w_n)$;
    \item serve as a basis for constructing efficienct exact designs either via rounding (e.g., Chapter 12 of \cite{pukelsheim}) applied to $\supp(\w^*_\infty)$ or through exact design computation restricted to $\supp(\w^*_\infty)$ or $\tilde{\supp}_\infty$;
    \item enable methods for computing optimal or nearly-optimal exact designs based on approximation by a surrogate criterion, for which fast discrete optimization algorithms are available (\cite{aqua1}, \cite{FilovaHarman20});
    \item  in “rational” models, allow construction of optimal exact designs of various sizes corresponding to the vertices of an optimal polytope (\cite{HFR24}).
\end{itemize}
This paper introduces yet another application of optimal approximate designs for constructing optimal exact designs: the removal of provably redundant candidates from $\X$. While $\supp(\w^*_\infty)$ for some $\w^*_\infty \in \Xi^*_\infty$, or more broadly $\tilde{\supp}_\infty$, can also be viewed as a reduction of the candidate set, these sets are not generally guaranteed to contain the support of an optimal size-$n$ design. In contrast, the method we propose ensures that the reduced candidate set retains this guarantee.

In this paper, we also assume that an arbitrary nonsingular size-$n$ design $\w^+_n$ is available. This design need not be optimal; although, in general, the better $\w^+_n$ is, the better the performance of our proposed methods. In practice, such a design is easily obtained. One simple option is to construct $\w^+_n$ quickly using any of the exact design algorithms mentioned above, applied only to the typically small sets $\supp(\w^*_\infty)$ or $\tilde{\supp}_\infty$. This completely circumvents running any optimal exact design algorithm on the potentially huge initial candidate set.

\subsection{Outline of the paper}

The purpose of this paper is to provide necessary conditions for each index that belongs to the support of an optimal size-$n$ design, that is, for any
\begin{equation*}
    \ell \in \supp^*_n:=\cup_{\w^*_n \in \Xi^*_n} \supp(\w^*_n).
\end{equation*}
These conditions can be verified numerically in a reliable manner, enabling a prescan of all candidates in $\mathcal{X}$ and the removal of those that do not satisfy them. From an optimization perspective, removing redundant candidates is equivalent to eliminating superfluous variables from the corresponding optimization problem. This reduction can make the search for an optimal size-$n$ design feasible in terms of memory requirements and can also substantially accelerate computation. In practical problems, the candidate set can often be reduced by several orders of magnitude. Although potentially valuable for any $D$-optimal exact design algorithm or heuristic, we specifically demonstrate that this preprocessing step can make mixed-integer second-order cone solvers applicable and effective even for massive initial design spaces. As an additional theoretical result, we prove that for sufficiently large $n$, the supports of optimal size-$n$ designs are subsets of a maximum-variance set $\tilde{\supp}_\infty$ based on an optimal approximate design.

Candidate-point elimination methods have already been proposed for $D$-optimal approximate designs (see \cite{del1}, \cite{del2}, and \cite{del3}). Here, we show that theorems of a similar type, albeit utilizing a different approach, also hold for the more challenging $D$-optimal exact design problem. To our knowledge, this is the first work that develops broadly applicable and efficiently computable necessary conditions for identifying redundant points in the exact optimal design setting.

The paper is organized as follows. Section~\ref{sec:Inequalities} introduces necessary conditions that must be satisfied by every support index of an optimal size-$n$ design. If $i \in [N]$ fails any of these conditions, candidate $\x_i$ can be removed from $\X$ without affecting optimality. Specifically, in Subsection~\ref{sec:relax}, we present a fast “augmentation” condition, related to earlier work such as branch-and-bound procedures for exact $D$-optimality. This condition is simple, computationally inexpensive, and often powerful. Moreover, it implies that $\tilde{\supp}_\infty$ covers the supports of all optimal size-$n$ designs for a sufficiently large $n$. Subsection~\ref{sec:exchange} then introduces a fundamentally different “exchange” condition, which is more computationally involved but can provide considerably greater reductions in the candidate set. Section~\ref{sec:Examples} demonstrates the approach with both simulated and real-world examples with candidate sets of sizes up to $N \approx 10^8$, and Section~\ref{sec:Conclusions} summarizes the main findings and outlines several promising directions for future research. The appendix provides an overview of the notation and the proofs of all lemmas.

\section{Conditions for optimal support}\label{sec:Inequalities}

\subsection{Augmentation condition}\label{sec:relax}

A straightforward necessary condition for an index $\ell$ to belong to the support of an optimal size-$n$ design is the following: a single trial at $\x_\ell$ must be augmentable by $n - 1$ additional trials to yield a design that is at least as good as a known size-$n$ design $\w^+_n$. Formally, if $\ell \in \supp^*_n$, then 
\begin{equation}\label{eq:augbnd}
       \max_{\w \in \Xi_n, w_\ell \geq 1/n} \Phi(\M(\w)) \geq \Phi(\M(\w^+_n)).
\end{equation}
Computing the maximum on the left-hand side of \eqref{eq:augbnd} is a discrete optimization problem of difficulty comparable to the original exact design problem. However, a wide range of upper bounds on this maximum can be constructed, yielding weaker, but numerically more tractable, necessary conditions for $\ell \in \supp^*_n$. 

In fact, this approach is the one typically used for pruning the nodes in branch-and-bound (BnB) algorithms. One way to construct an upper bound on the maximum in \eqref{eq:augbnd} is through direct algebraic methods; see, for example, formula (6) in \cite{Welch} based on the Hadamard inequality or the spectral bound from \cite{KLW}. However, our numerical experiments suggest that these bounds, while apparently effective within the BnB framework, perform relatively poorly for pruning the BnB nodes corresponding to the very weak constraints of the type $w_\ell \geq 1/n$, i.e., for removing redundant individual candidates. 
 
A simpler yet still powerful approach is to relax \eqref{eq:augbnd} by substituting $\Xi_\infty$ for $\Xi_n$, leading to the following necessary condition for $\ell \in \supp^*_n$:
\begin{equation}\label{eq:relbnd}
       \max_{\w \in \Xi_\infty, w_\ell \geq 1/n} \Phi(\M(\w)) \geq \Phi(\M(\w^+_n)).
\end{equation}
For smaller candidate sets, it may be possible to solve the convex problem  on the left-hand side of  \eqref{eq:relbnd} using an iterative convex optimization procedure. In the context of BnB algorithms for optimal exact design, a broad spectrum of convex algorithms has been used for this purpose (e.g., \cite{Welch}, \cite{UcinskiPatan2007},  \cite{HarmanSagnolBnB} or \cite{Ahipasaoglu2021}). However, for large candidate sets, solving a separate convex problem for each candidate is computationally too costly.

In contrast, our goal is to develop a general method specific to the problem of candidate reduction, one that remains computationally efficient even for very large candidate sets. To this end, we derive generally powerful necessary conditions for $\ell \in \supp^*_n$ that can be verified rapidly. 

First, observe that the feasible set $\{\w \in \Xi_\infty : w_\ell \geq 1/n\}$ in \eqref{eq:relbnd} is a polytope with vertices
\begin{equation}\label{eq:wils}
    \w_{i,\ell} := \frac{n-1}{n} \: \e_i + \frac{1}{n}\e_\ell
\end{equation}
for $i \in [N]$,  where $\e_i$ is the $i$-th standard basis vector. Any design $\w$ in this polytope can be expressed as a convex combination
\begin{equation}\label{eq:alphas}
    \w=\sum_{i \in [N]} \alpha_{i,\ell}^{\w} \w_{i,\ell}, \quad \alpha_{i,\ell}^{\w} :=
\begin{cases}
\dfrac{n w_i - 1}{n-1}, & i = \ell, \\[1.2ex]
\dfrac{n w_i}{n-1}, & i \neq \ell,
\end{cases}
\end{equation}
that is, $\alpha_{i,\ell}^{\w}$ are nonnegative numbers summing to $1$. This observation underpins a simple proof of the following lemma, as well as a bound used later in Section \ref{sec:exchange}. 
\begin{lemma}\label{lem:relbnd}
Let $\ell \in \supp^*_n$, $\w^+_n \in \Xi_n$, and let $\N$ be any $m \times m$ positive definite matrix. Then, 
\begin{multline}\label{eq:lem}
    \f^\top_\ell\N^{-1}\f_\ell \geq mn \frac{\Phi(\M(\w_n^+))}{\Phi(\N)}\\- (n-1)\max_{i \in [N]} \f^\top_i\N^{-1}\f_i.
\end{multline}
\end{lemma}

Lemma~\ref{lem:relbnd} is conceptually close to much of the research in convex optimization and is therefore related to several existing results. For instance, besides our original proof, an alternative proof can be derived from the construction of a dual solution to the optimal design problem. (Both proofs, along with all others, are given in the appendix.)
 
Any positive definite $\N$ in Lemma~\ref{lem:relbnd} provides a necessary condition for $\ell$ to support an optimal size-$n$ design. Hence, it is possible to search for a matrix $\N=\N_\ell$ that leads to the strongest necessary condition for $\ell \in \supp^*_n$. However, our numerical analysis suggests that the straightforward choice $\N := \M_*$, where $\w^*_\infty$ is an optimal approximate design, is a very good compromise between simplicity and power. This matrix is independent of $\ell$, typically available, and still effective in identifying a large number of candidates for removal. With this choice of $\N$, noting that $\Phi(\M(\w_n^+))/\Phi(\N)=\eff_\infty(\w^+_n)$ and using the equivalence theorem
$\max_i \f^\top_i \N^{-1} \f_i = \max_i v_i^* = m$,
we obtain: 
\begin{theorem}\label{thm:1}
    Let $\ell \in \supp^*_n$, $\w^+_n \in \Xi_n$, and let $\w^*_\infty \in \Xi^*_\infty$ be an optimal approximate design. Then,
    \begin{equation}\label{eq:1}
       v_\ell^* \geq mn \left(\eff_\infty(\w^+_n) - \frac{n-1}{n}\right).
    \end{equation}
\end{theorem}

If $\eff_\infty(\w^+_n) \leq (n-1)/n$, the right-hand side of \eqref{eq:1} is nonpositive, and the theorem provides no useful restriction. Nevertheless, Theorem \ref{thm:1} can be powerful if $\eff_\infty(\w^+_n)$ is close to $1$. In particular, if $\w_\infty^*$ is itself a size-$n$ design (as, e.g., in the model 3.1 from \cite{HFR24} or some models studied in \cite{Freise2020}), we can set $\w^+_n = \w_\infty^*$, and the condition of Theorem \ref{thm:1} reduces the candidate set to $\tilde{\supp}_\infty$. 

Moreover, when combined with Theorem 12.10 in \cite{pukelsheim}, Theorem~\ref{thm:1} yields the following result, which helps explain the widely observed tendency for the support of an optimal size-$n$ design to approach the support of optimal approximate designs as $n \to \infty$.
\begin{corollary}\label{cor:limit}
    There exists a natural number $n_0$ such that $\supp^*_n \subseteq \tilde{\supp}_\infty$ for all $n \geq n_0$.
\end{corollary}

\subsection{Exchange condition}\label{sec:exchange}

A substantially different idea leads to a generally stronger necessary condition. For an index $\ell$ to be in the support of an optimal size-$n$ design $\w^*_n$, no single-point exchange of $\x_\ell$ for another candidate $\x_i$ can be improving. Formally,
\begin{equation}\label{eq:exchange}
    \Phi\left(\M(\w^*_n) - \frac{1}{n}\f_\ell\f^\top_\ell + \frac{1}{n}\f_i\f_i^\top \right) \leq \Phi(\M(\w^*_n))
\end{equation}
for any $\ell \in \supp(\w^*_n)$ and any $i \in [N]$. A convenient alternative form of \eqref{eq:exchange}, obtained by applying, e.g., \cite[15.10(d)]{seber}, is
\begin{multline}\label{eq:trdet}
    n\tr\left(\M^{-1}(\w^*_n)[\f_i\f_i^\top - \f_\ell\f_\ell^\top]\right)\\
    \leq \det\left([\f_\ell \: | \: \f_i]^\top \M^{-1}(\w^*_n)[\f_\ell \: | \: \f_i]\right)
\end{multline}
for any $\ell \in \supp(\w^*_n)$ and any $i \in [N]$, where $[\f_\ell \: | \: \f_i]$ denotes the $m \times 2$ matrix with columns $\f_\ell$ and $\f_i$. We express \eqref{eq:trdet} via a 'standardized covariance matrix'
\begin{equation*}
    \B(\w_n^*):=\M^{1/2}_*\M^{-1}(\w^*_n)\M^{1/2}_*
\end{equation*}
and 'standardized rgressors' $\s_i:=\M_*^{-1/2}\f_i$, $i \in [N]$:
\begin{multline}\label{eq:trdetstd}
    n\tr\left(\B(\w^*_n)[\s_i\s_i^\top - \s_\ell\s_\ell^\top]\right)\\
    \leq \det\left([\s_\ell \: | \: \s_i]^\top \B(\w^*_n) [\s_\ell \: | \: \s_i]\right).
\end{multline} 

The subsequent strategy is to construct a lower bound on the trace on the left-hand side as well as an upper bound on the determinant on the right-hand side of \eqref{eq:trdetstd}.

\subsubsection{Trace and determinant bounds}

To compactly state a suitable lower bound on the trace in \eqref{eq:trdetstd}, define, for any pair of vectors $\v,\z \in \RR^m$, their 'discrepancy' as
\begin{equation}\label{eq:Delta}
    \Delta(\v,\z) := \frac{1}{2}\big(\lVert\v + \z\rVert \lVert \v - \z \rVert + \lVert\v\rVert^2 - \lVert\z\rVert^2\big),
\end{equation}
where $\|\cdot\|$ is the Euclidean norm of a vector. Note that $\Delta(\v,\z)$ is nonnegative and closely related to the distance between the matrices $\v\v^\top$ and $\z\z^\top$. Indeed, it is a matter of simple algebra to show that the spectral norm of the difference of the two matrices is
\begin{equation*}
\|\v\v^\top-\z\z^\top\|_2 =
\begin{cases}
\Delta(\v,\z), & \text{if } \|\v\| \geq \|\z\|, \\
\Delta(\z,\v), & \text{if } \|\z\| \geq \|\v\|.
\end{cases}
\end{equation*}

\begin{lemma}\label{lem:trdown}
    Let $\v,\z \in \RR^m$. Let $\B$ be a positive semidefinite $m \times m$ matrix and let $\gamma_1$ and $\gamma_m$ be the smallest and the largest eigenvalues of $\B$, respectively. Then
    \begin{equation*}
        \tr(\B[\v\v^\top - \z\z^\top]) \geq \gamma_1 \Delta(\v,\z) - \gamma_m \Delta(\z,\v).
    \end{equation*}
\end{lemma}

An appropriate upper bound on the determinant in \eqref{eq:trdetstd} is given by the following lemma.
\begin{lemma}\label{lem:detup}
    Let $\A$ be an $m \times k$ matrix, $k \in [m]$, and let $\B$ be a positive semidefinite $m \times m$ matrix. Let $\gamma_1\leq \cdots \leq \gamma_m$ be the eigenvalues of $\B$, repeated according to their multiplicities. Then
    \begin{equation*}
      \det(\A^\top \B \A) \leq \left(\prod\nolimits_{i=m-k+1}^m \gamma_i \right) \det(\A^\top \A).
    \end{equation*}
    In particular, for $k=2$ and $\A=[\v \: | \: \z]$, where $\v, \z \in \RR^m$, the bound simplifies to
    \begin{multline}\label{lem:detupk2}
     \det\left([\z \: | \: \v]^\top \B [\z \: | \: \v]\right)\\
     \leq \gamma_{m-1}\gamma_m (\|\v\|^2\|\z\|^2-(\v^\top \z)^2).   
    \end{multline}
\end{lemma}

Setting $\B=\B(\w_n^*)$, $\v=\s_i$, and $\z=\s_\ell$, Lemma~\ref{lem:trdown} and inequality~\eqref{lem:detupk2} from Lemma \ref{lem:detup}, together with~\eqref{eq:trdetstd}, yield the following necessary condition for $\ell \in\supp^*_n$:
\begin{multline}\label{eq:trdeteig}
    n\big(\gamma_1\Delta(\s_i,\s_\ell)-\gamma_m \Delta(\s_\ell,\s_i)\big)\\
    \leq \gamma_{m-1}\gamma_m\big(\|\s_i\|^2\|\s_\ell\|^2-(\s_i^\top \s_\ell)^2\big),
\end{multline}
where $\gamma_1$ is the smallest, $\gamma_{m-1}$ is the second largest, and $\gamma_m$ is the largest eigenvalue of $\B(\w^*_n)$. These eigenvalues are unknown; however, we can rapidly compute lower and upper bounds on the unknown eigenvalue-based quantities in \eqref{eq:trdeteig}, as we explain next.

\subsubsection{Eigenvalue bounds}

To construct eigenvalue bounds for $\B(\w^*_n)$, note first the following lower bound on the  determinant of $\B^{-1}(\w^*_n)$:
\begin{align} \label{eq:d}
    \det(\B^{-1}(\w^*_n)) &= \det(\M^{-1/2}_*\M(\w^*_n)\M^{-1/2}_*) \notag\\ 
    &\geq \det(\M^{-1/2}_*\M(\w^+_n)\M^{-1/2}_*) \notag\\
    &= \eff^m_\infty(\w^+_n) =:d_+.
\end{align}
In addition, utilizing \eqref{eq:wils} and \eqref{eq:alphas}, we obtain an upper bound on the trace of $\B^{-1}(\w^*_n)$ valid for any $\ell \in \supp(\w^*_n)$:
\begin{align}\label{eq:t}
   \tr(\B^{-1}(\w^*_n)) &= \tr\left[\sum_i \alpha_{i,\ell}^{\w^*_n} \M^{-1/2}_*\M(\w_{i,\ell})\M^{-1/2}_*\right] \notag\\
   &= \frac{n-1}{n} \sum_{i} \alpha_{i,\ell}^{\w^*_n} v_i^*+\frac{1}{n}v^*_\ell \notag\\
   &\leq \frac{n-1}{n} \max_{i} v_i^*+\frac{1}{n}v^*_\ell \notag\\
   &= \frac{n-1}{n} m + \frac{1}{n}v^*_\ell =: t_\ell.
\end{align}

Besides elementary matrix algebra, we used the equivalence theorem $\max_i v^*_i=m$. Note also that \eqref{eq:d}, the arithmetic mean - geometric mean inequality, and \eqref{eq:t} imply
\begin{align}\label{eq:inequaldt}
  d_+^{1/m} \leq \left(\det(\B^{-1}(\w^*_n))\right)^{1/m} \notag\\
  \leq \frac{1}{m}\tr(\B^{-1}(\w^*_n)) \leq \frac{t_\ell}{m}.
\end{align}
Therefore, any $\ell \in \supp^*_n$ satisfies $d_+^{1/m} \leq t_\ell/m$. A key observation is that the knowledge of $d_+$ and $t_\ell$ is enough to compute bounds on the (products of) eigenvalues of $\B(\w^*_n)$, as we show in the following lemma. 

\begin{lemma}\label{lem:ggg}
    Let $m \geq 2$ be fixed. Let $k \in [m-1]$ and $t>0$. Consider the function $R_{k,t}:[0,\frac{t}{k}] \to \RR$ defined by
\begin{equation}\label{eq:R}
      R_{k,t}(g):=\left[g^k\left(\frac{t-kg}{m-k}\right)^{m-k}\right]^{1/m}
\end{equation}
for $g \in [0,t/k]$. Then, $R_{k,t}(0)=R_{k,t}(\frac{t}{k})=0$, $R_{k,t}(\frac{t}{m})=\frac{t}{m}$, $R_{k,t}$ is strictly increasing on $[0,\frac{t}{m}]$, strictly decreasing on $[\frac{t}{m},\frac{t}{k}]$, and strictly concave on the entire $[0,\frac{t}{k}]$. Let $d>0$ be such that $d^{1/m} \leq \frac{t}{m}$. For the equation 
\begin{equation*}
  R_{k,t}(g)=d^{1/m},
\end{equation*}
let $\underline{g}(k,t,d)$ be the unique solution in the interval $[0,\frac{t}{m}]$ and $\overline{g}(k,t,d)$ the unique solution in the interval $[\frac{t}{m},\frac{t}{k}]$. Next, let $\B$ be an $m \times m$ positive definite matrix with eigenvalues $\gamma_1 \leq \cdots \leq \gamma_m$, such that $\det(\B^{-1}) \geq d$ and $\tr(\B^{-1}) \leq t$. Then, for any selection of indices $1\leq i_1 < \cdots < i_k \leq m$,
   \begin{equation}\label{eq:prodgam}
      \underline{g}\left(k, t, d\right) \leq \prod_{j=1}^k \gamma_{i_j}^{-1/k} \leq \overline{g}\left(k, t, d\right).
   \end{equation}
The above inequalities also hold for $k=m$ if we define $\underline{g}(m,t,d):=d^{1/m}$ and $\overline{g}(m,t,d):=t/m$.
\end{lemma}

Hence, \eqref{eq:d}, \eqref{eq:t} and \eqref{eq:inequaldt} imply that we can use Lemma \ref{lem:ggg} with $k=1,2$, $d=d_+$, $t=t_\ell$ and $\B=\B(\w^*_n)$ to obtain bounds for the (products of) eigenvalues of $\B(\w^*_n)$:
\begin{align}
    \gamma_1 \geq (\overline{g}(1,t_\ell,d_+))^{-1},\label{eq:zab0}\\
    \gamma_m \leq (\underline{g}(1,t_\ell,d_+))^{-1},\label{eq:zab1}\\
    \gamma_{m-1}\gamma_m \leq (\underline{g}(2,t_\ell,d_+))^{-2}.\label{eq:zab2}
\end{align}

Due to the favorable properties of $R_{k,t}(g)$, in particular its concavity, the numerical computation of $\overline{g}(k,t_\ell,d_+)$ and $\underline{g}(k,t_\ell,d_+)$ is essentially equivalent to having an analytic formula for these numbers. 

By combining \eqref{eq:trdeteig} and \eqref{eq:zab0}-\eqref{eq:zab2}, we obtain:
\begin{theorem}\label{thm:2}
Let $\ell \in \supp^*_n$ and let $\w^+_n \in \Xi_n$ be a nonsingular size-$n$ design. Let $\s_i=\M_*^{-1/2}\f_i$ for all $i \in [N]$, $d_+$ be given by \eqref{eq:d} and $t_\ell$ by \eqref{eq:t}. Then $d_+^{1/m} \leq \frac{1}{m}t_\ell$ and
    \begin{multline}\label{eq:3}
    \frac{\Delta(\s_i,\s_\ell)}{\overline{g}(1,t_\ell,d_+)} -\frac{\Delta(\s_\ell,\s_i)}{\underline{g}(1,t_\ell,d_+)} \\
    \leq \frac{\|\s_i\|^2\|\s_\ell\|^2-(\s_i^\top \s_\ell)^2 }{n\underline{g}^2(2,t_\ell,d_+)} 
    \end{multline}
    for all $i \in [N]$. 
\end{theorem}

The statement $d_+^{1/m} \leq \frac{1}{m}t_\ell$ in Theorem \ref{thm:2} is precisely the augmentation condition from Lemma \ref{lem:relbnd}, which follows from the definitions of $d_+$ and $t_\ell$ in \eqref{eq:d} and \eqref{eq:t}. Consequently, it is enough to verify \eqref{eq:3} only for those candidates that have already passed the simple check based on Theorem \ref{thm:1}. 

Moreover, for the numerical verification of \eqref{eq:3} by means of vector arithmetic, note that it can be expressed as 
\begin{multline}\label{eq:4}
   \min_{i \in [N]} \left\{\|\s_i\|^2\|\s_\ell\|^2-(\s_i^\top \s_\ell)^2 - q_\ell(\|\s_i\|^2 - \|\s_\ell\|^2) \right. \\
   + \left. r_\ell\sqrt{(\|\s_i\|^2 + \|\s_\ell\|^2)^2-4(\s_i^\top \s_\ell)^2}\right\} \geq 0,
\end{multline}
where the constants
\begin{eqnarray*}
   q_\ell &:=& \frac{n}{2}\underline{g}^2(2,t_\ell,d_+)\left(\frac{1}{\underline{g}(1,t_\ell,d_+)}+\frac{1}{\overline{g}(1,t_\ell,d_+)}\right),\\
   r_\ell &:=& \frac{n}{2}\underline{g}^2(2,t_\ell,d_+)\left(\frac{1}{\underline{g}(1,t_\ell,d_+)}-\frac{1}{\overline{g}(1,t_\ell,d_+)}\right)
\end{eqnarray*}
do not depend on $i$ and can be rapidly precomputed for each $\ell$ satisfying $d_+^{1/m} \leq \frac{1}{m}t_\ell$. 

An interesting observation is that the quantities in \eqref{eq:3} and \eqref{eq:4} can be expressed in terms of statistical characteristics of mean-response predictors based on $\w^*_\infty$, under the standard linear regression model $E(y)=\f^\top\boldsymbol{\theta}$, $\Var(y)\equiv \sigma^2$. Specifically,
\begin{align*}
    \|\s_i\|^2=v^*_i \propto \Var \left(\widehat{\f_i^\top\theta}\right),\\
    (\s_i^\top\s_\ell)^2 \propto \cov^2\left(\widehat{\f_i^\top\theta}, \widehat{\f_\ell^\top\theta}\right).
\end{align*} 
Similarly, the constants $q_\ell$ and $r_\ell$ are implicit functions of $v^*_\ell$ and $\eff_\infty(\w^+_n)$. However, developing a clear statistical interpretation of the exchange support conditions remains a challenge.
\bigskip

\begin{figure*}[ht]
    \centering
    \includegraphics[width=0.9\textwidth]{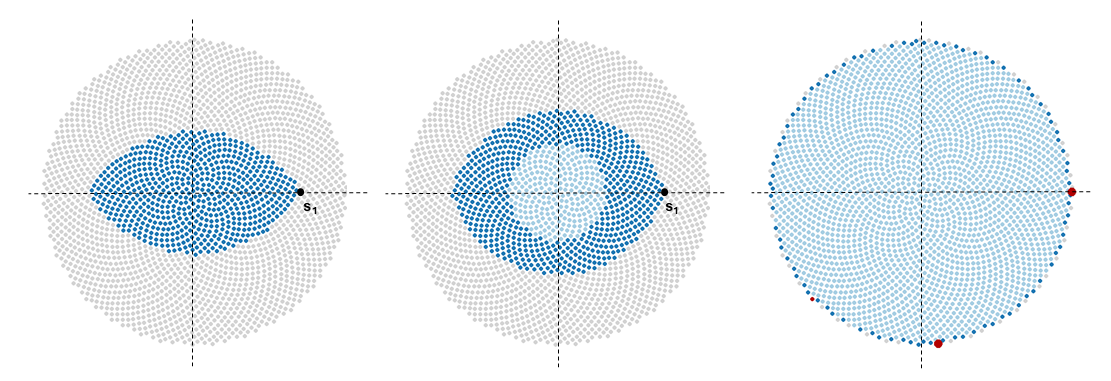} 
    \caption{Illustration of the removal methods based on Theorems \ref{thm:1} and \ref{thm:2}. We consider an artificial model with $m=2$, $n=9$, and $N=2500$ regressors: three fixed points $\mathbf{s}_1=(1,0)^\top$, $\mathbf{s}_2=(\sqrt{2},0)^\top$, $\mathbf{s}_3=(0,\sqrt{2})^\top$, and $2497$ Fibonacci points in the disk of radius $\sqrt{2}$ centered at $(0,0)^\top$. In this setting, $\M_*=\I_2$, i.e., $\s_i = \f_i$ for all $i \in [N]$. Panel (a): $\sqrt{d_+}=\eff_\infty(\w^+_9)=0.8$. At this efficiency, the augmentation condition from Theorem~\ref{thm:1} cannot eliminate any candidate because $\eff_\infty(\w^+_9)<(n-1)/n \approx 0.889$, so the right-hand side of \eqref{eq:1} is negative. By contrast, the exchange condition of Theorem~\ref{thm:2} (inequality~\eqref{eq:3}) is violated for many pairs $(\ell,i)$, enabling reduction. For instance, dark-blue points mark those $\s_\ell$ that violate \eqref{eq:3} with $i=1$. Evidently, even when $\eff_\infty(\mathbf{w}^+_n)$ is as low as $0.8$, the exchange condition of Theorem~\ref{thm:2}, applied by varying $i$ over all indices, would eliminate the vast majority of candidates. Panel (b): $\sqrt{d_+}=\eff_\infty(\w^+_9)=0.9$. Light-blue points are regressors eliminated by Theorem~\ref{thm:1}. The dark-blue points are regressors $\s_\ell$ not removed by Theorem~\ref{thm:1} but violating the exchange condition of Theorem~\ref{thm:2} for $i=1$. Panel (c): $\sqrt{d_+}=\eff_\infty(\w^+_9)=0.994$, the efficiency of an exact design $\w^+_9$ computed on $\supp(\w^*_\infty) = \{2,3\}$. Here, almost all points are light-blue, i.e., eliminated by Theorem~\ref{thm:1}. Dark-blue points are further removed by Theorem~\ref{thm:2} after checking \eqref{eq:3} for all $i$ (not just for $i=1$ as above). Gray and red points mark the $54$ remaining candidates which, by Theorem~\ref{thm:2}, are guaranteed to contain the support of any optimal exact design of size $6$. The true size-$9$ optimal design uses the regressors shown in red: the small red point represents a single trial, while the two larger red points each represent four trials.}
    \label{fig:circles}
\end{figure*}

Finally, to aid geometric intuition, we provide in Figure \ref{fig:circles} an illustration of both Theorems \ref{thm:1} and \ref{thm:2} for a simple problem with $m=2$, $n=9$, and $N=2500$.

\section{Numerical examples}\label{sec:Examples}

\subsection{Example 1: Gaussian regressors}\label{subsec:demo1}

For a triplet $(c,m,n)$ of integers, consider the procedure $G(c,m,n)$ that generates a size-$n$ design problem with random regressors $\f_1,\ldots,\f_{10^c} \sim \mathcal{N}_m(\0_m, \I_m)$.
To examine the numerical behavior of the proposed candidate removal methods, we used triplets $(6 \pm 2,5,35)$, $(6,5 \pm 2,35)$, and $(6,5,35 \pm 25)$, resulting in seven different settings. Note that the candidate set sizes $N=10^c$ for all considered $c=4,6,8$ are too large to directly apply the MISOCP implementations with available hardware.\footnote{We used a 64-bit Windows 11 system with an Intel Core i7-9750H processor (release 2019) operating at 2.60 GHz, with 16 GB of RAM. All codes required to reproduce the numerical results are available from the authors upon request.} For each triplet $(c,m,n)$, we repeated the following procedure $20$ times:
\begin{enumerate}
    \item[0.] Generate an optimal size-$n$ design problem via $G(c,m,n)$;
    \item[1.] Compute an optimal approximate design $\w^*_\infty$ using the \texttt{od\_REX} function of the \texttt{R} package \texttt{OptimalDesign} (\cite{rlib});\footnote{More precisely, we computed an approximate design that achieved a lower bound of $1-10^{-9}$ on the efficiency relative to the theoretical optimum; we consider such a design numerically perfectly optimal.}
    \item[2.] Compute a size-$n$ design $\mathbf{w}^+_n$ on the support of $\w^*_\infty$ using the MISOCP routine \texttt{od\_MISOCP} from \texttt{OptimalDesign}, run to optimality;\footnote{\texttt{OptimalDesign} performs the MISOCP computations with the \texttt{gurobi} solver under an academic license. We used the MIPGap set to $10^{-3}$.}    
    \item[3.] Remove redundant candidates based on $\w^*_\infty$ and $\w^+_n$ using the augmentation condition (Theorem~\ref{thm:1});
    \item[4.] Further remove redundant candidates using the exchange condition (Theorem~\ref{thm:2});
    \item[5.] Apply the MISOCP solver to the reduced candidate set until optimality is reached.
\end{enumerate}

We remark that after Step 1 all regressors $\f_i$ were linearly transformed by $\M^{-1/2}_*$. This transformation preserves the set of optimal size-$n$ designs, as it multiplies the determinant of every design by the same constant factor. It requires negligible time and improves both the speed and stability of subsequent computations, effectively serving as a preconditioning step. In addition, it provides the vectors $\s_i$ used in Theorem \ref{thm:2}. We also note that the support of $\w_\infty^*$ from Step 1 was smaller than $30$ for each $(c, m, n)$. The designs $\w_n^+$ in Step 2 were therefore easy to compute via MISOCP for such small candidate sets $\supp(\w^*_\infty)$.

\begin{figure*}[htbp]
    \centering
    \includegraphics[width=0.45\textwidth]{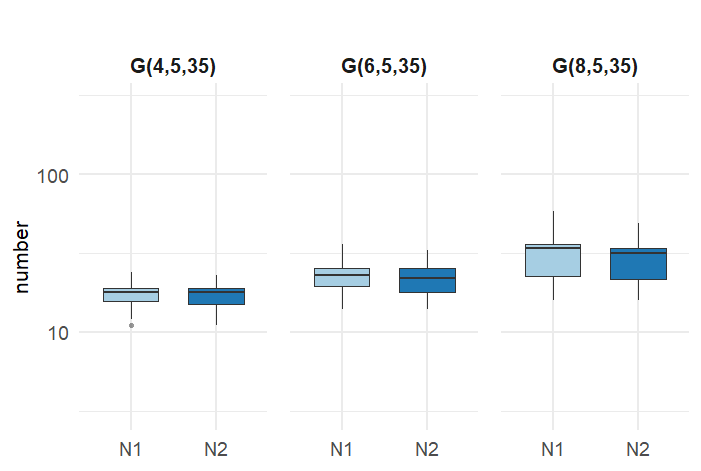}\hfill
    \includegraphics[width=0.45\textwidth]{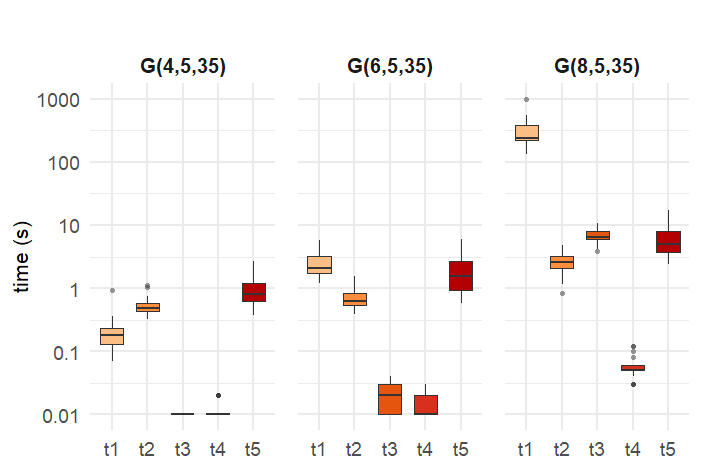}

    \vspace{1em} 

    \includegraphics[width=0.45\textwidth]{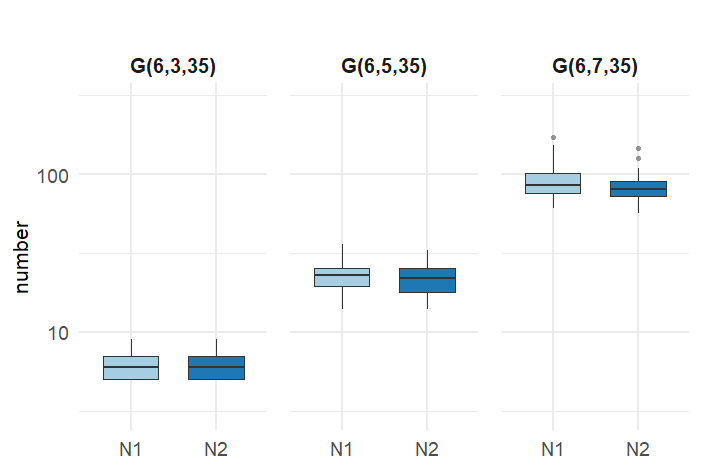}\hfill
    \includegraphics[width=0.45\textwidth]{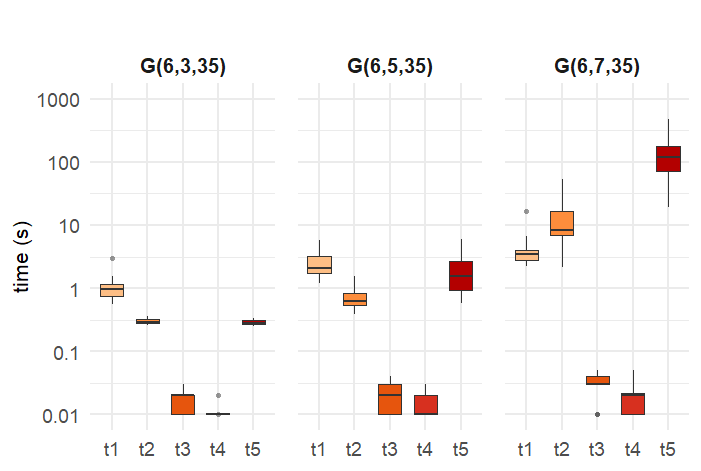}

    \vspace{1em} 

    \includegraphics[width=0.45\textwidth]{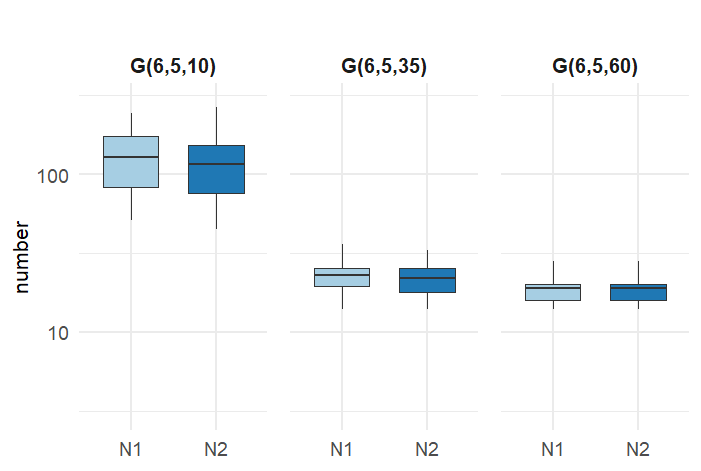}\hfill
    \includegraphics[width=0.45\textwidth]{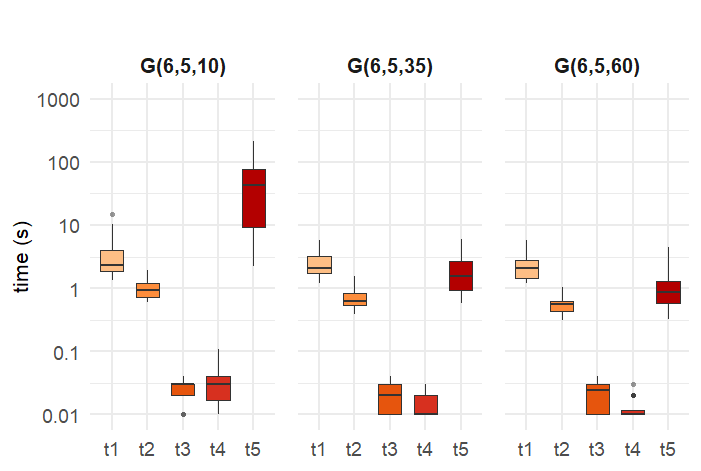}

    \caption{Summary of the numerical study in Section~\ref{subsec:demo1} for settings $G(c,m,n)$, where $N=10^c$ is the size of the initial candidate set, $m$ is the regressor dimension, and $n$ is the required size of the exact design. Left column: number of candidates remaining after the augmentation condition (Theorem~\ref{thm:1}, labeled N1) and after the exchange condition (Theorem~\ref{thm:2}, labeled N2). Right column: computation times of the five phases described in Section~\ref{subsec:demo1}: t1 - computation of $\mathbf{w}^*_\infty$; t2 - computation of $\mathbf{w}^+_n$; t3 - evaluation of the augmentation condition for all candidates; t4 - evaluation of the exchange condition for candidates not removed in Step 3; t5 - final MISOCP computation. Rows: top - $N$ varies; middle - $m$ varies; bottom - $n$ varies. All vertical axes are logarithmic.}
    \label{fig:Gauss}
\end{figure*}

The numerical results are shown in Figure~\ref{fig:Gauss}. In the left column of panels, we see that the augmentation condition alone drastically reduces the candidate set: from $N$ between $10^4$ and $10^8$ down to about $100$ or fewer candidates, with the strongest effect for small $m$ and large $n$. The exchange condition provides (in this class of models) only moderate additional reduction, becoming noticeable mainly for large $m$ and small $n$. The computation times are shown in the right column of panels in Figure~\ref{fig:Gauss}. Steps 1 and 2 usually finish in under $10$ seconds, except for computing $\w^*_\infty$ in Step 1 with $N=10^8$ candidates, which required several hundred seconds on average. Steps 3 and 4 are also typically very fast, with only a minor slowdown in Step 3 for the most demanding case $N=10^8$. Finally, Step 5 takes time that positively correlates with the number of remaining candidates, as expected.

The first panel in the left column of Figure~\ref{fig:Gauss} shows that the number of non-removed candidates depends only weakly on the size of the initial candidate set. This can be explained using known theoretical results: if $\ell$ lies in the support of any $D$-optimal approximate design, then $\f_\ell$ must be a vertex of the Elfving set (see Section 8.5 of \cite{pukelsheim}). Consequently, asymptotic bounds on the number of vertices of symmetric Gaussian polytopes (e.g., \cite{Hug2004}) provide upper bounds on the number of support points of $D$-optimal approximate designs with Gaussian regressors and, in the limit, also of the corresponding $D$-optimal exact designs (cf. Corollary \ref{cor:limit}). Combining Theorems 3.3 and 4.2 in \cite{Hug2004} shows that, for large $N$, the number of vertices of the symmetric Gaussian polytope grows no faster than $C_m (\log(N))^{(m-1)/2}$, i.e., at a rate slower than any positive polynomial power of $N$. This matches our empirical findings on the large magnitude of candidate set reduction for the Gaussian random regressors.

\subsection{Example 2: A mixture model}\label{subsec:demo2}

Although the Gaussian regressors from the previous example are convenient for exploring how the performance of an optimal design algorithm depends on the problem characteristics $N$, $m$, and $n$, they are not typical of real applications. By contrast, one of the most common practical settings for optimal design is the planning of mixture experiments (see, e.g., \cite{amtataw2025}, \cite{sharma2025}, \cite{khamtree2025} or \cite{paucar2025} for recent applications of $D$-optimal mixture designs).

\begin{figure*}[ht]
    \centering
    \includegraphics[width=0.90\textwidth]{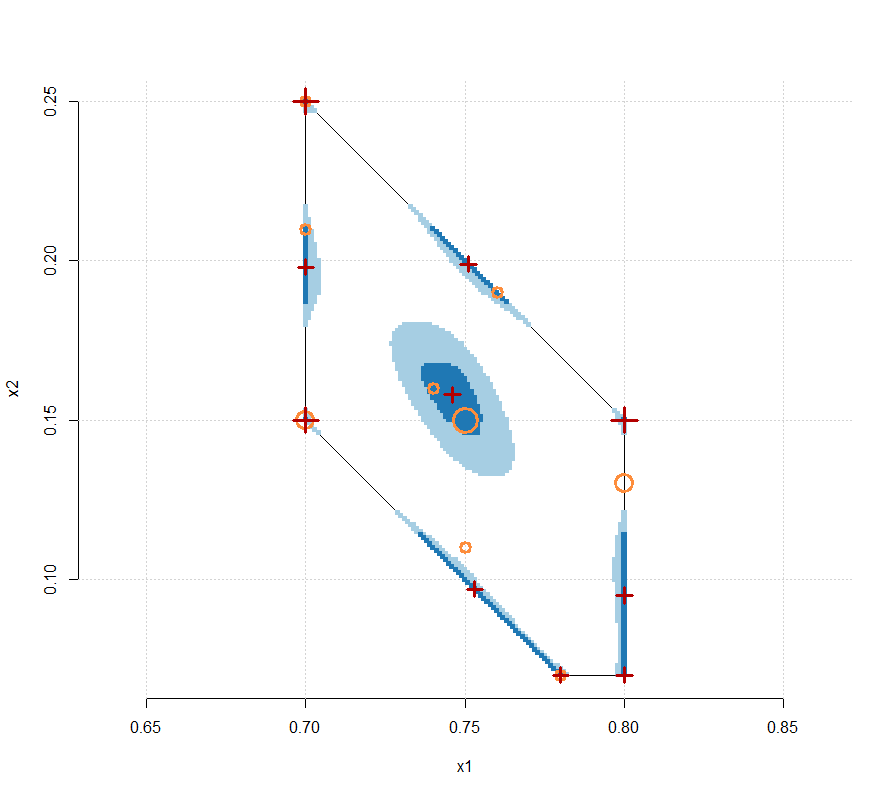} 
    \caption{Projection $\mathbf{x}=(x_1,x_2,x_3)^\top \mapsto (x_1,x_2)^\top$ of the candidate set for the Scheffé mixture model with interval constraints on the three mixture proportions; See Section \ref{subsec:demo2}. Orange circles denote the original design, and red pluses the design computed via MISOCP by utilizing our candidate removal method; both designs use $n=13$ trials. Symbol sizes indicate replications: three sizes of circles correspond to one, two, or three replications, while two sizes of pluses correspond to one or two replications. Candidates remaining after reduction by the augmentation condition (Theorem~\ref{thm:1}) are shown in light blue, and those remaining after further reduction by the exchange condition (Theorem~\ref{thm:2}) in dark blue.}
    \label{fig:MIX}
\end{figure*}

Therefore, we adopted a mixture experimental design problem from \cite{paucar2025}, in which the researchers used three mixture components and a quadratic Scheffé mixture model with $m = 6$ parameters. The candidate set, a projection of which is shown in Figure \ref{fig:MIX}, forms a two-dimensional pentagon in $\RR^3$ due to factor-level constraints: the mixture components must sum to one, and each is restricted to a relatively narrow predefined interval. For such practically common yet irregular candidate spaces, no ready-made experimental designs are available, so an efficient design must be computed numerically. The design reported in the referenced paper is given to two decimal places (i.e., integer percentage points), and the size-13 design used by the experimenters is depicted in Figure \ref{fig:MIX} by orange circles; its $D$-criterion value is $1.169 \times 10^{-4}$.

For this setting, we computed the $D$-optimal design on a discretization of the design space to three decimal places. Specifically, we considered the set $\tilde{\X}$ of vectors $(x_1, x_2, x_3)^\top$ satisfying $x_1 \in [0.7, 0.8]$, $x_2 \in [0.07, 0.25]$, $x_3 \in [0.05, 0.15]$, $x_1 + x_2 + x_3 = 1$, and such that $10^3x_1$, $10^3x_2$, and $10^3x_3$ are integers. The $N = 9991 \approx 10^4$ elements of $\tilde{\X}$ were indexed as $\x_1, \ldots, \x_N$, and, in accordance with the quadratic Scheffé model, we defined the regressors $\f_i=(x_{i1},x_{i2},x_{i3}, x_{i1}x_{i2}, x_{i1}x_{i3}, x_{i2}x_{i3})^\top$, $i \in [N]$, where $x_{ij}$ is the $j$-th component of $\x_i$. That is, $m=6$.

To demonstrate the effectiveness of the removal methods, we applied the same procedure as in Section \ref{subsec:demo1}, except that in Step 5, the MISOCP procedure was terminated after $60$ seconds of computation. The approximate computation times of Steps 1 to 4 were: $0.18$, $1.36$, $0.01$, and $0.49$ seconds. Figure \ref{fig:MIX} shows the $N_1 = 1644$ candidates remaining after the application of the augmentation condition and the $N_2 = 390$ points remaining after also applying the exchange condition. Hence, in this example, the additional reduction from the exchange condition is substantial. 

While the size of the initial optimization problem exceeds available memory and cannot be solved directly, the removal of redundant candidates made it feasible to run the MISOCP solver. After 60 seconds, the computation reached a gap of $0.87\%$ and returned a design with a $D$-criterion value of $1.495 \times 10^{-4}$, shown in Figure~\ref{fig:MIX} by red plus symbols. Compared with the design from \cite{paucar2025}, the new design shifts the placement of several points and improves the $D$-criterion value for the quadratic Scheffé mixture model. Nevertheless, the original design remains fairly efficient and includes three replications at the central point, which, while slightly reducing statistical efficiency, may offer practical advantages for variance estimation.

\begin{figure*}[htbp]
    \centering
    \includegraphics[width=0.48\textwidth]{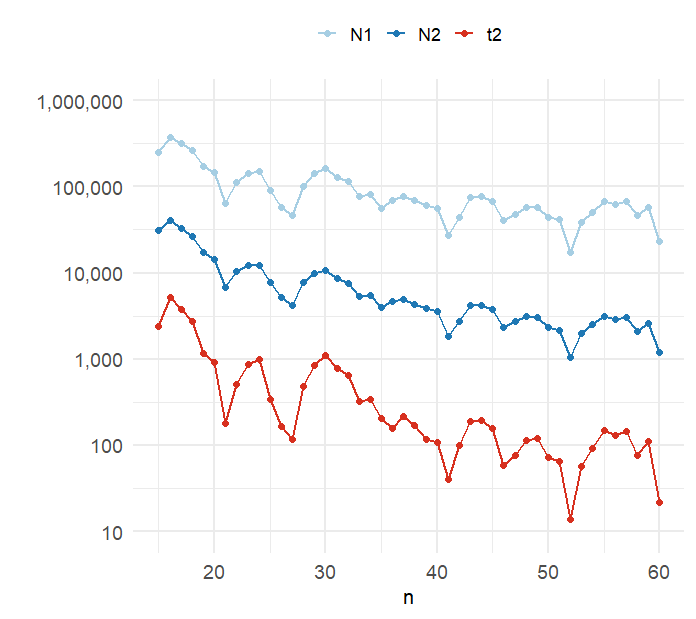}\hfill
    \includegraphics[width=0.48\textwidth]{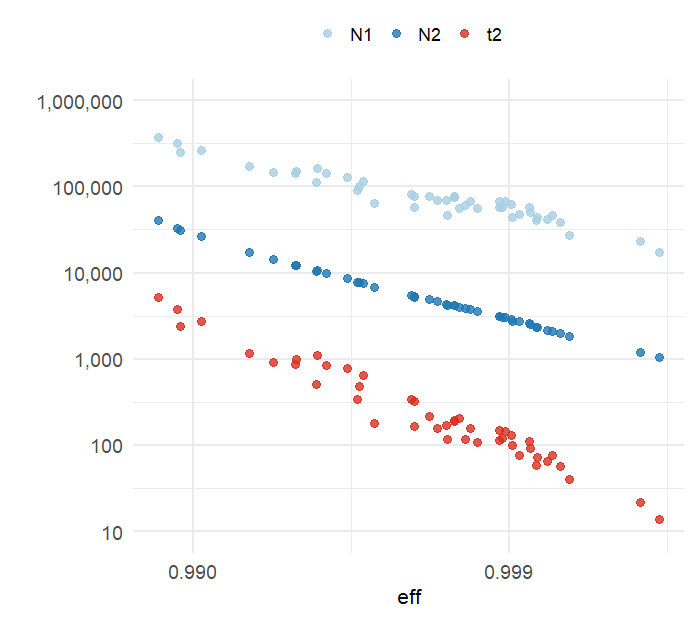}
    \caption{Summary of the second numerical study in Section~\ref{subsec:demo2} with $N \approx 10^6$ initial candidate points. Left panel: the number of remaining candidates after the augmentation condition (light blue, N1), after the exchange condition (dark blue, N2), and the computation time (in seconds) for the exchange step (red, t2) as functions of the design size $n$. Right panel: dependence of the same quantities on the efficiency $\eff_\infty(\w^+_n)$. See Section~\ref{subsec:demo2} for discussion of the main conclusions. The common vertical axis of both panels is logarithmic. The horizontal axis in the right panel uses a logarithmic transformation that makes efficiency values $0.99$, $0.999$, $0.9999$, etc., equally spaced.}
    \label{fig:mixture}
\end{figure*}

Removal of redundant candidate points based on Theorems~\ref{thm:1} and~\ref{thm:2} can also be applied to a grid arising from the requirement that factor levels be specified with $4$ decimal places, yielding $N = 981901 \approx 10^6$ candidates. We use this setting to highlight aspects of the proposed removal methods that do not arise in the random models of Section~\ref{subsec:demo1}. The left panel of Figure~\ref{fig:mixture} shows how the extent of candidate removal depends on the  sample size $n$. A first important observation is that for this problem, the exchange condition provides a considerable additional reduction beyond the augmentation condition - more than a tenfold decrease in candidate set size. At the same time, although the number of retained candidates generally decreases with $n$, the relationship is highly non-monotonic. The cause of this non-monotonicity is that for certain values of $n$, the combinatorial structure allows the $n$ trials of $\w^+_n$ to approximate the proportions of the optimal approximate design $\w^*_\infty$ more closely, while for others it does not, which in turn strongly affects the extent of candidate removal. Thus, compared to $n$, a better predictor of the reduction magnitude is the efficiency of $\w^+_n$ relative to $\w^*_\infty$. For the number $N_2$ of candidates retained after applying the exchange condition, we observed an almost perfectly linear dependence of $\log(N_2)$ on $\log(1 - \eff_\infty(\w^+_n))$; see the right panel of Figure~\ref{fig:mixture}. 

\section{Conclusions}\label{sec:Conclusions}

The augmentation condition (Theorem~\ref{thm:1}) is simple to state, straightforward to implement, and often has a large impact. Although we demonstrated its utility only for the MISOCP algorithm, it can serve as a natural preprocessing step for any $D$-optimal exact design algorithm or heuristic, provided that an optimal approximate design is available. In practice, this assumption is not restrictive: with modern algorithms, an optimal approximate design can typically be computed to essentially maximal possible numerical precision even for very large candidate sets, and such a design is also useful in several other ways. Moreover, even for cases where a perfectly optimal approximate design is not available, Lemma~\ref{lem:relbnd} offers an alternative.

The exchange condition (Theorem~\ref{thm:2}) can substantially further reduce the candidate set in some problems. It can be crucial in cases where the augmentation condition alone does not produce a sufficiently small set, and our experience suggests that it is particularly effective when the candidate set arises from a practically relevant discretization of a continuous design space. However, the usefulness of the exchange condition should be assessed on a case-by-case basis, since it is somewhat more complex to implement and more computationally demanding.

From a broader perspective, Corollary~\ref{cor:limit} suggests that the reduced design space tends to shrink as $n$ increases, approaching the support of the optimal approximate designs. This provides further justification for the common practice of basing the computation of efficient size-$n$ designs, if $n$ is large, on the optimal approximate design. It may also help explain the seemingly paradoxical observation that optimal exact design problems with larger sample sizes $n$ can, in some cases, be easier to solve using exact integer or mixed-integer programming methods.

Candidate removal is valuable not only for practical computation but also for research into $D$-optimal exact design algorithms on finite design spaces. As we have seen, the size of the initial candidate set is often a poor proxy for the true difficulty of the problem; a more meaningful measure is the number of non-removable candidates. For example, models with Gaussian regressors tend to be relatively easy, whereas models used in practice, in particular with many candidates $i$ satisfying $v^*_i \approx m$, may be much more challenging.

A limitation of the proposed method is that in certain highly symmetric models, no candidate reduction is possible: any candidate can be mapped to any other by a criterion-preserving transformation; therefore, all candidates support some optimal size-$n$ design. However, these cases are largely well-studied settings for which tabulated or theoretical optimal designs are available, so computation is often not needed. By contrast, in the problems that typically motivate computation, such as models with regressors of random nature, asymmetric candidate spaces, or nonlinear models linearized in an estimated nominal parameter, major candidate-set reduction may be both feasible and highly useful.

This work opens up several promising directions for further research. One extension would be to adapt the proposed approach to models with multivariate responses and/or additional constraints restricting the exact designs $\w_n$ from the discretized simplex $\Xi_n$ to complex subsets of $\Xi_n$. Another important avenue is to develop candidate-removal methods for other optimality criteria, notably for $A$-optimality. Promising research directions also lie at the interface with theoretical computer science, for instance in assessing the computational complexity of exactly determining the set $\supp^*_n$.

\backmatter


\section*{Declarations}

\begin{itemize}
\item Funding. This work was supported by the Slovak Scientific Grant Agency (VEGA), Grant 1/0480/26.
\item Competing interests. The authors have no relevant financial or non-financial interests to disclose.
\item Ethics approval and consent to participate. Not applicable.
\item Consent for publication. Not applicable.
\item Data availability. Not applicable.
\item Materials availability. Not applicable.
\item Code availability. All codes used for the numerical results are available upon request from the authors.
\item Author contribution. Both coauthors contributed to the conceptualization, methodology, mathematical proofs, software, writing and reviewing of the article.
\end{itemize}


\begin{appendices}

\section{Proofs} 

\noindent \textbf{Proof of Lemma \ref{lem:relbnd}}

\begin{proof} The gradient of $\Phi$ at $\N$ is $\nabla\Phi(\N)=m^{-1}\Phi(\N)\N^{-1}$ (e.g., page 79 of \cite{pukelsheim}), therefore, noting that there exists an optimal size-$n$ design $\w_n^*$ with the $\ell$-th component at least $1/n$ expressible in the form \eqref{eq:alphas}, concavity of $\Phi$ and elementary algebra give
\begin{align*}
    \Phi(\M(\w_n^+)) &\leq \Phi(\M(\w_n^*)) \\
    & \leq \Phi(\N) + \tr\left[\nabla\Phi(\N)(\M(\w_n^*)-\N)\right] \\
    &= \frac{1}{m} \Phi(\N) \: \tr\left[\N^{-1}\M\left(\sum_{i}\alpha^{\w_n^*}_{i,\ell}\w_{i,\ell}\right)\right]\\
    &= \frac{1}{m} \Phi(\N) \sum_{i}\alpha^{\w_n^*}_{i,\ell} \tr\left[\N^{-1}\M\left(\w_{i,\ell}\right)\right]\\
    &\leq \frac{1}{m} \Phi(\N) \max_{i} \tr\left[\N^{-1}\M\left(\w_{i,\ell}\right)\right].
\end{align*}
Combining this inequality and the definition of $\w_{i,\ell}$, we obtain \eqref{eq:lem}. 
\end{proof}

\noindent \textbf{An alternative proof of Lemma \ref{lem:relbnd}}

\begin{proof}
For brevity, let $\Xi^{\ell}_\infty$ denote the convex set $\{\w \in \Xi_\infty \;|\; w_\ell \geq 1/n\}$.

\cite{madan} observed that a feasible dual solution to the optimal design problem can be constructed from exact designs. They employed this dual solution to establish efficiency bounds for exact designs and, ultimately, to prove approximation guarantees for the Fedorov exchange algorithm. Here, we adopt a similar approach to construct dual solutions for the $D$-optimality problem over $\Xi^{\ell}_\infty$. Unlike in \cite{madan}, however, our aim is to use this construction to derive Lemma \ref{lem:relbnd}.
 
According to \cite{Ahipasaoglu2021}, the dual problem corresponding to
\begin{equation}\label{eq:Primal}
       \max_{\w \in \Xi^{\ell}_\infty} \ln\Phi(\M(\w))
\end{equation}
can be written as
\begin{align}\label{eq:Dual}
	\min_{\H, \lambda_1,\ldots,\lambda_N} & -m^{-1} \ln\det(\H) \\
	s.t. \quad & \f_i^\top\H\f_i = m + \lambda_\ell/n - \lambda_i, \: i \in [N], \notag \\
    & \lambda_1,\ldots,\lambda_N \geq 0, \notag\\
    & \H \text{ is positive definite.}
\end{align}

Let $\N$ be an $m \times m$ positive definite matrix. Then, a feasible solution for \eqref{eq:Dual} is 
\begin{equation}\label{eq:DualFeasible}
    \H = c \N^{-1}, \: \lambda_i = \frac{nm - cv_\ell}{n-1}  - c v_i, \: i \in [N],
\end{equation}
where
\begin{equation*}
  c := \frac{nm}{v_\ell + (n-1)v_{\max}} \: \text{ and } \: 
  v_i := \f_i^\top \N^{-1}\f_i,\: i \in [N].
\end{equation*}
Indeed, let $i \in [N]$. Observe that $v_i = \f_i^\top \N^{-1} \f_i = c^{-1} \f_i^\top \H \f_i$.
It is straightforward to verify that
\begin{equation*}
    \begin{aligned}
\frac{\lambda_\ell}{n} - \lambda_i 
&= \frac{1}{n}\left(\frac{nm - cv_\ell}{n-1}  - c v_\ell\right) - \frac{nm - cv_\ell}{n-1} + c v_i \\
&= cv_i - m = \f_i^\top \H \f_i - m,
\end{aligned}
\end{equation*}
which shows that the equality constraints \eqref{eq:Dual} are satisfied. However, the inequality constraints in \eqref{eq:Dual} are also satisfied, since $v_i \leq v_{\max}: = \max_{i \in [N]} \f_i^\top \N^{-1}\f_i$ yields
\begin{equation*}
    \begin{aligned}
\lambda_i &\geq \frac{nm}{n-1} - c\left(\frac{v_\ell}{n-1} + v_{\max}\right) \\
&= \frac{nm}{n-1} - \frac{nm}{v_\ell + (n-1)v_{\max}} \frac{v_\ell + (n-1)v_{\max}}{n-1} =0.
\end{aligned}
\end{equation*}

Let $\w \in \Xi^{\ell}_\infty$. Then, $\w$ is feasible for \eqref{eq:Primal} and $(\H,  \lambda_1,\ldots,\lambda_N)$ is feasible for \eqref{eq:Dual}. Weak duality then gives
\begin{equation*}
\ln\Phi(\M(\w)) \leq -m^{-1} \ln\det(\H),
\end{equation*}
that is
\begin{align*}
\Phi(\M(\w)) &\leq (\det(\H))^{-1/m} \\
             &=  (\det(c\N^{-1}))^{-1/m} = c^{-1} (\det(\N))^{1/m}.
\end{align*}
Therefore
\begin{align*}
\Phi(\M(\w^+_n)) &\leq \max_{\w \in \Xi^{\ell}_\infty} \Phi(\M(\w)) \\
                 &\leq \left(\frac{nm}{v_\ell + (n-1)v_{\max}}\right)^{-1} \Phi(\N),
\end{align*}
which is equivalent to \eqref{eq:lem}.
\end{proof}

\noindent \textbf{Proof of Corollary \ref{cor:limit}}

\begin{proof} Theorem 12.10 in \cite{pukelsheim} implies that for some $h>0$ and any $n$ there exists a size-$n$ design $\w_n^+$ satisfying $\eff_\infty(\w_n^+) \geq 1 - \frac{h}{n^2}$. The definition of $\tilde{\supp}_\infty$ implies that there exists a positive $\varepsilon$ smaller than $m - \max_{i \notin \tilde{\supp}_\infty}v_i^*$. Define $n_0:=\lceil mh/\varepsilon \rceil$. Then, for all $n \geq n_0$ and any $i \notin \tilde{\supp}_\infty$, we have
\begin{align*}
    v_i^* &< m-\varepsilon = m \left(1 - \frac{h}{n_0}\right) \leq m\left(1 - \frac{h}{n}\right) \\
    &= mn\left(1-\frac{h}{n^2}-\frac{n-1}{n}\right) \\
    &\leq mn\left(\eff_\infty(\w_n^+) - \frac{n-1}{n}\right).
\end{align*}
In view of Theorem \ref{thm:1}, such an $i$ cannot be in the support of any optimal size-$n$ design.
\end{proof}

\noindent \textbf{Proof of Lemma \ref{lem:trdown}}

\begin{proof}
The statement of the lemma is straightforward to verify if $\v$ and $\z$ are collinear. Assume the opposite. Let $\C:=\v\v^\top-\z\z^\top$. As $\rank(\C)=2$ and $\C=\G_1\G_2$, where $\G_1 = [\v \: | \: \z]$ and $\G_2=[\v \: | \: -\z]^\top$, the (negative) minimum eigenvalue $\lambda_1$ of $\C$ and the (positive) maximum eigenvalue $\lambda_m$ of $\C$ are the same as those of
\begin{equation*}
\G_2\G_1 = 
\begin{pmatrix}
    \lVert\v\rVert^2 & \v^\top\z \\ -\z^\top\v & -\lVert\z\rVert^2
\end{pmatrix} =:\A.    
\end{equation*}
The eigenvalues of the $2 \times 2$ matrix $\A$ are just 
\begin{equation*}
 \begin{aligned}
& \frac{\tr(\A)}{2} \pm \left(\frac{\tr^2(\A)}{4} - \det(\A) \right)^{1/2} = \frac{\lVert\v\rVert^2 - \lVert\z\rVert^2}{2} \\
& \pm \left[\frac{(\lVert\v\rVert^2 - \lVert\z\rVert^2)^2 - 4((\v^\top\z)^2-\lVert\v\rVert^2\lVert\z\rVert^2)}{4} \right]^{1/2} \\
&=\frac{1}{2}\left(\lVert\v\rVert^2 - \lVert\z\rVert^2\pm 
\lVert\v + \z\rVert \lVert \v - \z \rVert\right),
\end{aligned}
\end{equation*}
and these two values are exactly $\Delta(\v,\z)$ and $-\Delta(\z,\v)$, respectively.

Consider the spectral decomposition of the rank-two matrix $\C$: $\C = \lambda_m\u_m\u_m^\top + \lambda_1\u_1\u_1^\top$, where $\u_1, \u_m$ are the orthonormal eigenvectors corresponding to $\lambda_1, \lambda_m$. Applying standard algebra, we see that
\begin{equation*}
  \tr(\B\C) = \lambda_m \u_m^\top \B \u_m + \lambda_1 \u_1^\top \B \u_1 
            \geq \lambda_m \gamma_1 + \lambda_1 \gamma_m.
\end{equation*}
\end{proof}

\noindent \textbf{Proof of Lemma \ref{lem:detup}}

\begin{proof}
Let $\supp_{m,k}$ denote the system of all $k$-element subsets of $[m]$. For a $k \times m$ matrix $\G$ and $I \in \supp_{m,k}$, let $\G_I$ be the submatrix of $\G$ formed by the columns of $\G$ with indices from $I$. Let $\B=\U\Gamma\U^\top$, where $\U$ is an orthogonal matrix and $\Gamma=\diag(\gamma_1,\ldots,\gamma_m)$. Denote $p:=\prod_{i=m-k+1}^m \gamma_i$ and use the Cauchy-Binet formula to obtain
\begin{align*}
    \det(\A^\top \B\A) &= \det(\A^\top \U \Gamma \U^\top \A) \\
    &={\sum}_{I \in \supp_{m,k}}{\det}^2((\A^\top \U \Gamma^{1/2})_I) \\
    &= {\sum}_{I \in \supp_{m,k}}\det(\Gamma_I){\det}^2((\A^\top \U)_I) \\
    &\leq p {\sum}_{I \in \supp_{m,k}}{\det}^2((\A^\top \U)_I) \\
    &= p \det(\A^\top \U\U^\top \A) = p \det(\A^\top \A).
\end{align*}
\end{proof}

\noindent \textbf{Proof of Lemma \ref{lem:ggg}}

\begin{proof} 
Consider the assumptions of the lemma. The properties of the function $R_{k,t}$ can be verified using the standard techniques of calculus. Proving \eqref{eq:prodgam} is also elementary, but we exhibit a fast proof. Let $p:=\prod_{i=1}^k \gamma_i^{-1}$ and $s:=\sum_{i=k+1}^m \gamma_i^{-1}$. The arithmetic mean - geometric mean (AM-GM) inequality implies $kp^{1/k} + s \leq (\sum_{i=1}^k \gamma_i^{-1}) + s \leq t$, that is, $t-kp^{1/k} \geq s$, hence
\begin{align}
  p\left(\frac{t-kp^{1/k}}{m-k}\right)^{m-k} &\geq p\left(\frac{s}{m-k}\right)^{m-k} \notag \\
  &\geq p \prod_{i=k+1}^m \gamma_i^{-1} \geq d. \label{eq:ps}
\end{align}
Also note that the AM-GM inequality yields
\begin{align*}
    p^{1/k} &=\prod_{i=1}^k \gamma^{-1/k} \leq \frac{1}{k}\sum_{i=1}^k \gamma^{-1} \\
    &\leq \frac{1}{k}\sum_{i=1}^m \gamma^{-1} \leq \frac{t}{m}.
\end{align*}
Since $p^{1/k} \in [0,\frac{t}{k}]$, we see that \eqref{eq:ps} implies $R_{k,t}(p^{1/k}) \geq d^{1/m}$, and the properties of $R_{k,t}$ mean that $p^{1/k}$ must lie between the two roots of $R_{k,t}$, i.e.,
   \begin{equation*}
      \underline{g}\left(k, t, d\right) \leq p^{1/k} \leq \overline{g}\left(k, t, d\right).
   \end{equation*}
We proved the required inequality for $i_j=j$, $j=1,\ldots,k$. The inequalities for general $1 \leq i_1<\cdots <i_k \leq m$ follow by symmetry.
\end{proof}

\section{Notation} 

\begin{itemize}[label={},leftmargin=30pt,align=parleft, labelsep=0.5em,labelwidth=3.2em]
\item[$\RR$] Set of real numbers \item[$\mathbb{N}$] Set of natural numbers $\{1,2,\ldots\}$ \item[$\lbrack k\rbrack$] Set $\{1,\ldots,k\}$ (where $k\in\mathbb{N}$) \item[$N$] Initial number of candidate points \item[$m$] Dimension of regressors and information matrices \item[$n$] Sample size; required size of the exact design \item[$\0_m$] Zero vector of size $m$ \item[$\e_i$] The $i$-th standard basis vector (with implicitly defined size) \item[$\x_i$] The $i$-th candidate point \item[$\f_i$] The regressor of dimension $m$ corresponding to $\x_i$ \item[$\s_i$] Standardized regressor $\M_*^{-1/2}\f_i$ \item[$\w$] Design vector $(w_1,\ldots,w_N)^\top$ \item[$\w^*_\infty$] A $D$-optimal approximate design \item[$\w^*_n$] A $D$-optimal exact design of size $n$ \item[$\w^+_n$] A nonsingular exact design of size $n$ \item[$\X$] Initial set of candidate points $\{\x_1,\ldots,\x_N\}$ \item[$\Xi_n$] Set of all exact designs of size $n$ \item[$\Xi_\infty$] Set of all approximate designs \item[$\Xi^*_n$] Set of all $D$-optimal exact designs of size $n$ \item[$\Xi^*_\infty$] Set of all $D$-optimal approximate designs \item[$\supp(\w)$] Support of design $\w$; that is, $\{i: w_i>0\}$ \item[$\supp^*_n$] Union of the supports of all $D$-optimal size-$n$ designs \item[$\supp^*_\infty$] Union of the supports of all $D$-optimal approximate designs \item[$\tilde{\supp}_\infty$] Maximum-variance set $\{i: v_i^*=m\}$ \item[$\I_m$] Identity matrix of size $m\times m$ \item[$\M(\w)$] Information matrix $\sum_{i\in[N]} w_i \f_i\f_i^\top$ \item[$\M_*$] Information matrix of a $D$-optimal approximate design $\w^*_\infty$ \item[$\B(\w^*_n)$] Standardized covariance matrix, that is, $\B(\w^*_n)=\M_*^{1/2}\M^{-1}(\w^*_n)\M_*^{1/2}$ \item[$\top$] Transpose of a matrix or vector \item[$\tr$] Matrix trace \item[$\det$] Matrix determinant \item[$\diag$] Diagonal matrix formed from given entries \item[$\rank$] Rank of a matrix \item[$\Phi(\M)$] $D$-optimality criterion $(\det(\M))^{1/m}$ \item[$v_i(\w)$] Variance function $\f_i^\top \M^{-1}(\w)\f_i$ \item[$v_i^*$] Variance function of an optimal approximate design, that is, $v^*_i=v_i(\w^*_\infty)$ \item[$\eff_\infty(\w)$] Efficiency of $\w$ relative to $\w^*_\infty$ \item[$\Delta(\v,\z)$] Discrepancy between $\v$ and $\z$; see \eqref{eq:Delta} \item[$R_{k,t}(g)$] Auxiliary function yielding bounds on eigenvalues; see \eqref{eq:R}
\end{itemize}

\end{appendices}


\end{document}